\documentclass[12pt]{article}
\usepackage{amssymb,amsmath,epsfig}
\allowdisplaybreaks

\begin{document}

\title{\bf Thermodynamics of Gravitationally Induced Particle Creation Scenario in DGP Braneworld}
\author{\textbf{Abdul Jawad} \thanks {jawadab181@yahoo.com; abduljawad@ciitlahore.edu.pk},
\textbf{Shamaila Rani} \thanks {shamailatoor.math@yahoo.com} and
\textbf{Salman Rafique} \thanks{salmanmath004@gmail.com}\\
Department of Mathematics, COMSAT Institute of\\ Information
Technology, Lahore-54000, Pakistan.}

\date{}
\maketitle
\begin{abstract}
In this paper, we discuss the thermodynamical analysis for
gravitationally induced particle creation scenario in the framework
of DGP braneworld model. For this purpose, we consider apparent
horizon as the boundary of the universe. We take three types of
entropy such as Bakenstein entropy, logarithmic corrected entropy
and power law corrected entropy with ordinary creation rate
$\Gamma$. We analyze the first law and generalized second law of
thermodynamics analytically for these entropies which hold under
some constraints. The behavior of total entropy in each case is also
discussed which implies the validity of generalized second law of
thermodynamics. Also, we check the thermodynamical equilibrium
condition for two phases of creation rate, that is constant and
variable $\Gamma$ and found its validity in all cases of entropy.
\end{abstract}
\textbf{Keywords:} Gravitationally induced particle creation; Thermodynamics;
Entropy corrections; DGP Brane-world.\\
\textbf{PACS:} 95.36.+d; 98.80.-k.

\section{Introduction}

It is believed that the universe undergoes an accelerated expansion
due to mysterious form of force called dark energy (DE) which was
firstly predicted by two independent teams of cosmologists
\cite{ref1,ref2}. Both of them used distant type Ia supernova as
standard candles to measure the expansion of the universe. This
discovery was unexpected, because before this invention,
cosmologists just think that the expansion of the universe would be
decelerating because of the gravitational attraction of the matter
in the universe. In the accelerated expansion of the universe, DE
plays major role but its nature is still unknown. The simplest
candidate of DE is the cosmological constant, but its composition
and mechanism are unknown. More generally, the detail of its
equation of state (EoS) and relationship with the standard model of
particle physics continue to be investigated both through
observationally and theoretically \cite{1b}. In order to explain
this cosmic acceleration, various DE models and modified theories of
gravity have been developed such as $f(R)$ theory, $f(T)$ theory,
Brans-Dicke theory, dynamical Chern-Simons modified gravity, DGP
Braneworld model, etc.

The concept of thermodynamics in cosmological system originates
through black hole physics. It was suggested \cite{10J} that the
temperature of Hawking radiations emitting from black holes is
proportional to their corresponding surface gravity on the event
horizon. Jacobson \cite{11J} found a relation between thermodynamics
and the Einstein field equations. He derived this relation on the
basis of entropy-horizon area proportionality relation along with
first law of thermodynamics (also called Clausius relation)
$dQ=TdS$, where $dQ,~T$ and $dS$ indicate the exchange in energy,
temperature and entropy change for a given system. It was shown that
the field equations for any spherically symmetric spacetime can be
expressed as $TdS=dE+PdV$ ($E,~P$ and $V$ represent the internal
energy, pressure and volume of the spherical system) for any horizon
\cite{12J}.

The generalized second law of thermodynamics (GSLT) has been studied
extensively in the scenario of expanding behavior of the universe.
The GSLT states that \textit{the entropy of matter inside the
horizon plus entropy of the horizon remains positive and increases
with the passage of time} \cite{13J}. In order to discuss GSLT,
horizon entropy of the universe can be taken as one quarter of its
horizon area \cite{14J} or power law corrected \cite{15J} or
logarithmic corrected \cite{16J} forms. Many people have explored
the validity of GSLT of different systems including interaction of
two  fluid components like DE and dark matter \cite{17J}, as well as
interaction of  three components of fluid \cite{18J} in the FRW
universe by using simple horizon entropy of the universe. The
thermodynamical analysis widely performed in modified theories of
gravity \cite{sj}.

The gravitationally induced particle creation is another well-known
mechanism which was firstly introduced by Schrodinger \cite{S1} on
microscopic level. This mechanism was further extended by Parker et
al. towards quantum field theory in curved spacetimes
\cite{S2}-\cite{S6}. The macroscopic description of particle
creation mechanism induced by gravitational field was presented by
Prigogine et al. \cite{S7}. Later on, the covariant description of
this mechanism was developed \cite{S8,S9} as well as the physical
difference between particle creation and bulk viscosity was also
given \cite{S10}. The particle creation process can be described
with the inclusion of backreaction term in the Einstein field
equations whose negative pressure may help in explaining the cosmic
acceleration. In this way, various phenomenological models of
particle creation have been presented \cite{S11}-\cite{S16}. It is
also shown that phenomenological particle production
\cite{S17}-\cite{S20} help in explaining the cosmic acceleration and
paved the alternative way to the concordance $\Lambda$CDM model.

The fact that we resides in a three-dimensional space embedded in an
extra-dimensional world and five-dimensional models in which
universe would be a hypersurface has attain a great attention. The
four-dimensional Einstein's equations projected onto the brane have
been explored by Shiromizu et al. \cite{B6}. The approaches which
made on the basis of brane-world in the early-time cosmology favor a
particular model of cosmic evolution featured by quadratic relations
between the energy density and the Hubble parameter, dubbed
quadratic cosmology \cite{B7}.

Recently, by assuming the gravitationally induced particle scenario
with constant specific entropy and arbitrary particle creation rate
($\Gamma$), thermodynamics on the apparent horizon for FRW universe
has been discussed \cite{1j}. They have investigated the first law,
GSLT and thermodynamical equilibrium by assuming the EoS for perfect
fluid and put forward various constraints on $\Gamma$ for which
thermodynamical laws hold. Our aim is to discuss the thermodynamical
analysis on the apparent horizon for gravitationally induced
particle creation scenario with ordinary creation rate $\Gamma$ by
assuming entropies (Bakenstein entropy or usual entropy, logarithmic
corrected entropy and power law corrected entropy) in DGP braneworld
model. The scheme of the paper is as follows: In the next section,
we will present the basic equations of DGP brane-world, particle
creation rate and cosmological parameters. Section \textbf{3, 4, 5}
contain the discussion of thermodynamic quantities as well as its
laws corresponding to usual, logarithmic and power law corrected
entropies, respectively. The last section comprises of concluding
remarks on our results.

\section{Basic Equations}

A most particular version was proposed by Dvali et al. \cite{1}, in
which the four dimensional FRW universe is enclosed in a five
dimensional Minkoski bulk with infinite size. The gravitational laws
were obtained by adding an Einstein-Hilbert term to the action of
brane computed with the brane's intrinsic curvature. The presence of
such a term in the action is generically induced by quantum
corrections coming from the bulk gravity and its coupling with
matter living on the brane and must be included for a large class of
theories for self-consistency \cite{1d,1e}. Here, we consider
$3$-brane embedded in a $5D$ space-time with an intrinsic curvature
term included in the brane whose action can be written as
\begin{equation}\label{S35}
S_{\textmd{(5)}}=-\frac{1}{2\kappa^2}\int
d^5X\sqrt{-\tilde{g}}\tilde{R}+\int d^5X\sqrt{-\tilde{g}}L_m,
\end{equation}
where $L_m$ is the brane curvature term, given by
\begin{equation}\label{S38}
L_m=-\frac{1}{2\mu^2}\int d^4x\sqrt{-g}R,
\end{equation}
and $\kappa^2=8\pi G_{\textmd{(5)}}=M^{-3}_{\textmd{(5)}},~
\mu^2=8\pi G_{\textmd{(4)}}=M^{-2}_{\textmd{(4)}}$. The
Eq.(\ref{S35}) represents the Einstein-Hilbert action in five
dimensions for a five-dimensional metric $\tilde{g}{_{\textmd{AB}}}$
(bulk metric) of Ricci scalar $\tilde{R}$. Similarly, Eq.(\ref{S38})
indicates the Einstein-Hilbert action for the induced metric
$\tilde{g}{_{\textmd{cd}}}$ on the brane with $R$ appeared as its
scalar curvature. From  Eq.(\ref{S35}), we can get modified
Friedmann equation as \cite{2}
\begin{equation}\label{S}
H^2+\frac{k}{a^2}=\bigg(\sqrt{\frac{\rho}{3M^2_p}+\frac{1}{4r^2_c}}+\frac{\epsilon}{2r_c}\bigg)^2,
\end{equation}
where $H=\frac{\dot{a}}{a}$ is Hubble parameter with $a(t)$ is the
scale factor. Also, $\rho=\rho_m+\rho_D$, the subscripts $m$ and $D$
represent the energy densities corresponding to dark matter and DE
respectively, $r_c$ is the crossover length which represents the
scale that has length away from which gravity starts opening into
the bulk \cite{2}. Moreover it is the distance scale follow the
comparison among $4D$ and $5D$ effects of gravity and can be written
as \cite{2}
\begin{eqnarray}\label{S2}
r_{\textmd{c}}=\frac{M^2_p}{2M^3_5},
\end{eqnarray}
where $M^2_{\textmd{5}}$ stands for the $5D$ Planck mass and
$M^2_{\textmd{p}}$ is the $4D$ Planck mass.

For the spatially flat DGP braneworld $(k=0)$, Eq.(\ref{S}) reduces
to
\begin{eqnarray}\label{C1}
H^2-\frac{\epsilon}{r_{\textmd{c}}}H=\frac{\rho}{3M^2_p}.
\end{eqnarray}
There exist two different branches for the DGP model depending on
the sign of $\epsilon$. These are as follows:
\begin{itemize}
  \item For $\epsilon=+1$, there is a de Sitter solution for Eq.(\ref{C1})
with constant Hubble parameter, i.e., $H=\frac{1}{r_c}\Rightarrow
a(t)=a_0e^{\frac{t}{r_c}}$ in the absence of any kind of energy or
matter field on the brane (i.e., $\rho=0$). However, this branch
faces some problems like ghost instabilities \cite{4}.
  \item For $\epsilon = -1$, the accelerated expansion of the universe can
only be explained through the inclusion of DE component in the DGP
scenario.
\end{itemize}
We consider the latter case in the present work. The equation of
continuity for this model will become
\begin{equation}\label{F1+}
\dot{\rho}+\Theta\big(\rho+P+\Pi\big)=0,
\end{equation}
where $\Pi$ is a particle creation pressure which represents the
gravitationally induced process of particle creation and $\Theta=3H$
is the fluid expansion. Differentiating Eq.(\ref{C1}) and replacing
the value of $\dot{\rho}$ using Eq.(\ref{F1+}), we obtain
\begin{eqnarray}\label{T1}
\dot{H}=\frac{-H\big(\rho+p+\Pi\big)}{M^2_p(2H-\frac{\epsilon}{r_c})}.
\end{eqnarray}
The respective EoS for this model is giving by
$p=\big(\gamma-1\big)\rho$ with $\frac{2}{3}\leq\gamma\leq 2$. The
non-conservation of the total rate of change of number of particles,
$N=na^3$ with comoving volume $a^3$ and $n$ is the number density of
particle production in an open thermodynamical system yields
\begin{equation}\label{S3}
\dot{n}+\Theta n=n\Gamma,
\end{equation}
where $\Gamma$ is a particle creation rate has negative and positive
phases. Negative $\Gamma$ represents the particle destruction and
positive $\Gamma$ describes the elimination of particles.
Furthermore, a non-zero $\Gamma$ produces effective bulk viscous
pressure \cite{6}-\cite{12}.

Now using the Eqs.(\ref{F1+}), (\ref{S3}) and Gibbs relation, we get
\begin{eqnarray}\label{S4}
Tds=d\bigg(\frac{\rho}{n}\bigg)+pd\bigg(\frac{1}{n}\bigg).
\end{eqnarray}
An equation related to  the creation pressure $\Pi$ and the creation
rate $\Gamma$ has the form
\begin{eqnarray}\label{S5}
\Pi=-\frac{\Gamma}{\Theta}\big(\rho+p\big).
\end{eqnarray}
Under traditional assumption that the specific entropy of each
particle is constant, i.e., the process is adiabatic or isentropic.
This implies a dissipative fluid is similar to a perfect fluid with
a non-conserved particle number. To discuss cosmological parameters,
we insert Eqs.(\ref{S5}) and $p=(\gamma-1)\rho$ in Eq.(\ref{T1}), to
obtain
\begin{eqnarray}\label{S6}
\frac{\dot{H}}{H^2}=-\frac{3\gamma\big(H-\frac{\epsilon}{r_{\textmd{c}}}\big)
\big(1-\frac{\Gamma}{3H}\big)}{(2H-\frac{\epsilon}{r_c})}.
\end{eqnarray}
The deceleration parameter $q$ can be written as
\begin{eqnarray}\label{S8}
q=-\frac{\dot{H}}{H^2}-1=\frac{3\gamma(H-\frac{\epsilon}{r_c})
\big(1-\frac{\Gamma}{3H}\big)}{(2H-\frac{\epsilon}{r_c})}-1.
\end{eqnarray}
The effective EoS parameter for this model turns out to be
\begin{eqnarray}\label{S36}
\omega_{\textmd{eff}}=\frac{p+\Pi}{\rho}=\gamma\big(1-\frac{\Gamma}{3H}\big)-1.
\end{eqnarray}
This parameter has ability to explain the different phases of the
universe on the basis of $\Gamma$, i.e., if $\Gamma<3H$, then we
have quintessence era of the universe ($\omega_{\textmd{eff}}<-1$),
if $\Gamma>3H$ then effective EoS parameter represents the phantom
era of the universe ($\omega_{\textmd{eff}}>-1$) while for
$\Gamma=3H$, effective EoS parameter exhbits the cosmological
constant ($\omega_{\textmd{eff}}=-1$).

In the following, we will discuss first and second thermodynamical
laws in the presence of particle creation rate $\Gamma$ on the
apparent horizon.

\section{Thermodynamical Analysis with Usual Entropy}

For flat FRW universe, Hubble parameter coincides with the apparent
horizon as $R_{\textmd{A}}=\frac{1}{H}$. Differentiating the
apparent horizon with respect to time, we get
\begin{eqnarray}\label{S11}
\dot{R_A}=-\frac{\dot{H}}{H^2}
=\frac{3\gamma(H-\frac{\epsilon}{r_c})\big(1-\frac{\Gamma}{3H}\big)}
{(2H-\frac{\epsilon}{r_c})}.
\end{eqnarray}
The Bekenstein entropy and Hawking temperature of the apparent
horizon are given by $(8\pi=G=1)$
\begin{eqnarray}\label{S14}
S_{\textmd{A}}=\frac{A}{4} =\frac{R^2_A}{8}\quad \text{and} \quad
T_{\textmd{A}}=\frac{1}{2\pi R_A} =\frac{4}{R_A},
\end{eqnarray}
where $A=4\pi R^2_A$. The first law of thermodynamics at the horizon
can be obtained through the Clausius relation as
\begin{eqnarray}\label{S17}
-dE_{\textmd{A}}=T_{\textmd{A}}dS_{\textmd{A}}.
\end{eqnarray}
For the sake of convenance, we consider
$\Xi=T_{\textmd{A}}dS_{\textmd{A}}+dE_{\textmd{A}}$. The
differential $dE_{\textmd{A}}$ is the amount of energy crossing the
apparent horizon can be evaluated as \cite{13}
\begin{eqnarray}\label{S18}
-dE_{\textmd{A}}=\frac{1}{2}R^3_{\textmd{A}}(\rho+p)H dt
=\frac{3\gamma\big(H-\frac{\epsilon}{r_{\textmd{c}}}\big)}{2H}dt\label{S18}.
\end{eqnarray}
From Eq.(\ref{S14}), the differential of surface entropy at apparent
horizon yields
\begin{eqnarray}\label{S45}
dS_A=\frac{3\gamma}{4H}\frac{(H-\frac{\epsilon}{r_c})(1-\frac{\Gamma}{3H})}{(2H-\frac{\epsilon}{r_c})},
\end{eqnarray}
which leads to
\begin{eqnarray}\label{S19}
T_{\textmd{A}}dS_{\textmd{A}}=\frac{3\gamma\big(H-\frac{\epsilon}{r_{\textmd{c}}}\big)\big(1-\frac{\Gamma}{3H}\big)}
{\big(2H-\frac{\epsilon}{r_{\textmd{c}}}\big)}dt.
\end{eqnarray}
Thus, $\Xi$ turns out to be
\begin{eqnarray}\label{S43}
\Xi=3\gamma\big(H-\frac{\epsilon}{r}\big)\bigg(\frac{1-\frac{\Gamma}{3H}}{2H-\frac{\epsilon}{r_c}}-\frac{1}{2H}\bigg).
\end{eqnarray}
From this relation, it can be seen that first law of thermodynamics
holds (i.e., $\Xi\rightarrow0$) at the apparent horizon for
$\Gamma=3H\big(1-\frac{2H-\frac{\epsilon}{r_c}}{2H}\big)$.

Next, we will discuss the GSLT and thermodynamical equilibrium of a
system containing perfect fluid distribution bounded by apparent
horizon in DGG brane-world scenario. For GSLT, the total entropy of
the system can not be decrease, i.e.,
$d(S_{\textmd{A}}+S_{\textmd{f}})\geq0$. In this relation,
$S_{\textmd{A}}$ and $S_{\textmd{f}}$ appear as the entropy at
apparent horizon and the entropy of cosmic fluid enclosed within the
horizon, respectively. The Gibbs equation is given by
\begin{eqnarray}\label{S20}
T_{\textmd{f}}dS_{\textmd{f}}=dE_{\textmd{f}}+p dV,
\end{eqnarray}
where $ T_{\textmd{f}} $ is the temperature of the cosmic fluid and
$ E_{\textmd{f}}$ is the energy of the fluid ($E_{\textmd{f}}=\rho
V) $. The evolution equation for fluid temperature having constant
entropy can be described as \cite{14}
\begin{eqnarray}\label{S21}
\frac{\dot{T_{\textmd{f}}}}{T_{\textmd{f}}}=(\Gamma-\Theta)\frac{\partial
p}{\partial \rho}.
\end{eqnarray}
Eq.(\ref{S6}) leads to $\Gamma-\Theta=
\frac{\dot{H}(2H-\frac{\epsilon}{r_{\textmd{c}}})}{\gamma
H(H-\frac{\epsilon}{r_{\textmd{c}}})} $ which inserting in
Eq.(\ref{S21}) gives the following equation
\begin{eqnarray}\label{S22}
\ln\bigg(\frac{T_{\textmd{f}}}{T_{\textmd{0}}}\bigg)=\frac{2(\gamma-1)}{\gamma}\int\frac{dH}{H}
~~\Rightarrow~~
T_{\textmd{f}}=T_{\textmd{0}}\bigg(H^2-\frac{\epsilon}{r_{\textmd{c}}}H\bigg)^\frac{(\gamma-1)}{\gamma},
\end{eqnarray}
where $T_0$ is the constant of integration. The differential of the
fluid entropy can be obtained by using the Eq.(\ref{S20}) as follows
\begin{eqnarray}\nonumber
dS_{\textmd{f}}&=&-\frac{3\gamma
T^{-1}_{0}}{(2H-\frac{\epsilon}{r_{\textmd{c}}})}\big(H-\frac{\epsilon}{r_{\textmd{c}}}\big)\big
(1-\frac{\Gamma}{3H}\big)\big(H^2-\frac{\epsilon}{r_{\textmd{c}}}H\big)^\frac{1-\gamma}{\gamma}\\
&\times&\bigg(1-\frac{1}{2Hr_{\textmd{c}}}-\frac{3\gamma}{2H}\big(H-\frac{\epsilon}{r_{\textmd{c}}}\big)\bigg)dt\label{S23}.
\end{eqnarray}
Using Eqs.(\ref{S45}) and (\ref{S23}), we get the rate of change of
total entropy as
\begin{eqnarray}\nonumber
\dot{S}_{T}&=&\frac{3\gamma\big(H-\frac{\epsilon}{r_{\textmd{c}}}\big)
\big(1-\frac{\Gamma}{3H}\big)}{4H\big(2H-\frac{\epsilon}{r_{\textmd{c}}}\big)}\bigg[1+4T^{-1}_0H
\big(H^2-\frac{\epsilon}{r_{\textmd{c}}}H\big)^{\frac{1}{\gamma}-1}\\\label{S24}
&\times&\bigg(\frac{1}{2H}\frac{\epsilon}{r_{\textmd{c}}}-\frac{3\gamma}{2H}
\big(H-\frac{\epsilon}{r_{\textmd{c}}}\big)-1\bigg)\bigg],
\end{eqnarray}
where $S_{T}=S_{\textmd{A}}+S_{\textmd{f}}$. We discuss the validity
of GSLT on the basis of $\Gamma$ such that
\begin{itemize}
  \item \underline{$\Gamma< 3H$:} The GSLT holds if the following
  constraint
  \begin{eqnarray}\nonumber
1>4T^{-1}_0
\big(H^2-\frac{\epsilon}{r_{\textmd{c}}}H\big)^{\frac{1}{\gamma}-1}\bigg(H-\frac{\epsilon}{2r_{\textmd{c}}}+\frac{3\gamma}{2}
\big(H-\frac{\epsilon}{r_{\textmd{c}}}\big)\bigg)
\end{eqnarray}
satisfies. This shows that the GSLT satisfies in the qunitessence
era of the evolving universe.
  \item \underline{$\Gamma> 3H$:} For this case, we have the
  constraint
  \begin{eqnarray}\nonumber
1<4T^{-1}_0
\big(H^2-\frac{\epsilon}{r_{\textmd{c}}}H\big)^{\frac{1}{\gamma}-1}\bigg(H-\frac{\epsilon}{2r_{\textmd{c}}}+\frac{3\gamma}{2}
\big(H-\frac{\epsilon}{r_{\textmd{c}}}\big)\bigg),
\end{eqnarray}
which implies the GSLT holds in phantom era of the universe.
  \item \underline{$\Gamma= 3H$:} This case implies $\dot{S}_{T}=0$ in the cosmological constant era.
\end{itemize}

Replacing $dt$ to $\frac{dH}{\dot{H}}$ and integrating
Eq.(\ref{S24}), we get
\begin{eqnarray}\nonumber
S_{\textmd{T}}&=&S_{\textmd{A}}+S_{\textmd{f}}\nonumber\\
&=&-\frac{1}{8H^2\epsilon^2r_cT_0(1-\gamma)}\big(1-\frac{\epsilon}{H^2r_c}\big)^{\frac{-1}{\gamma}}
\big(1-\frac{H^2r_c}{\epsilon}\big)^{{\frac{-1}{\gamma}}}
\bigg[6\gamma^2\epsilon^2~_2F1(\frac{-1+\gamma}{\gamma}\nonumber\\
&,&-\frac{1}{\gamma}, 2-\frac{1}{\gamma},
\frac{\epsilon}{H^2r_c})(H^2-\frac{\epsilon}{r_c})^{\frac{1}{\gamma}}
(1-\frac{H^2r_c}{\epsilon})^{\frac{1}{\gamma}}+2(H^2-\frac{\epsilon}{r_c})^{\frac{1}{\gamma}}
r^2_c\bigg\{-4H\epsilon\nonumber\\
&\times&(-1+\gamma)~_2F1(-\frac{1}{2}, -\frac{1}{\gamma},
\frac{1}{2},
\frac{H^2r_c}{\epsilon},)(1-\frac{\epsilon}{H^2r_c})^{\frac{1}{\gamma}}+\gamma\bigg(6H^3(-1+\gamma)\nonumber\\
&\times&~_2F1(\frac{1}{2},
\frac{-1+\gamma}{\gamma}\frac{3}{2},\frac{H^2r_c}{\epsilon})(1-\frac{\epsilon}{H^2r_c})^{\frac{1}{\gamma}}+\epsilon^2~_2F1(\frac{-1+\gamma}{\gamma},
-\frac{1}{\gamma}, 2-\frac{1}{\gamma}
\nonumber\\
&,&\frac{\epsilon}{H^2r_c})(1-\frac{H^2r_c}{\epsilon})^{\frac{1}{\gamma}}
\bigg)\bigg\}+2H^2r^3_c(H^2-\frac{\epsilon}{r_c})^{\frac{1}{\gamma}}(-1+\gamma)\bigg(4H~_2F1(\frac{1}{2}
\nonumber\\
&,&\frac{-1+\gamma}{\gamma}, \frac{3}{2},
\frac{H^2r_c}{\epsilon})(1-\frac{\epsilon}{H^2r_c})^{\frac{1}{\gamma}}+\gamma\epsilon\bigg(-~_2F1(-\frac{1}{\gamma},
-\frac{1}{\gamma}, \frac{-1+\gamma}{\gamma},
\frac{\epsilon}{H^2r_c})\nonumber\\
&\times&(1-\frac{H^2r_c}{\epsilon})^{\frac{1}{\gamma}}+(1-\frac{\epsilon}{H^2r_c})^{\frac{1}{\gamma}}
(-1+(1-\frac{H^2r_c}{\epsilon})^{\frac{1}{\gamma}})\bigg)\bigg)+(-1+\gamma)\epsilon\nonumber\\
&\times& r_c\bigg\{6\gamma
H(H^2-\frac{\epsilon}{r_c})^{\frac{1}{\gamma}}\bigg(-2(1-\frac{\epsilon}{H^2r_c})^{\frac{1}{\gamma}}~_2F1(-\frac{1}{2},
\frac{-1}{\gamma}, \frac{1}{2},
\frac{H^2r_c}{\epsilon})-H\gamma\nonumber\\
&\times&(1-\frac{H^2r_c}{\epsilon})^{\frac{1}{\gamma}}~_2F1(-\frac{1}{\gamma},
-\frac{1}{\gamma}, \frac{-1+\gamma}{\gamma},
\frac{\epsilon}{H^2r_c})+(1-\frac{\epsilon}{H^2r_c})^{\frac{1}{\gamma}}\nonumber\\
&\times&(-1+(1-\frac{H^2r_c}{\epsilon})^{\frac{1}{\gamma}}\bigg)
-\epsilon(1-\frac{\epsilon}{H^2r_c})^{\frac{1}{\gamma}}(1-\frac{H^2r_c}{\epsilon})^{\frac{1}{\gamma}}\bigg\}\bigg].
\end{eqnarray}
The plot between total entropy and parameter $\gamma$ is shown in
Figure \textbf{1} for three values of $T$ by setting constant values
as $H=67$, $r_c=\frac{1}{67}$, $\epsilon=-1$. We observe that
$S_T\geq0$ for all the values of $T$ which leads to the validity of
GSLT.
\begin{figure} \centering
\epsfig{file=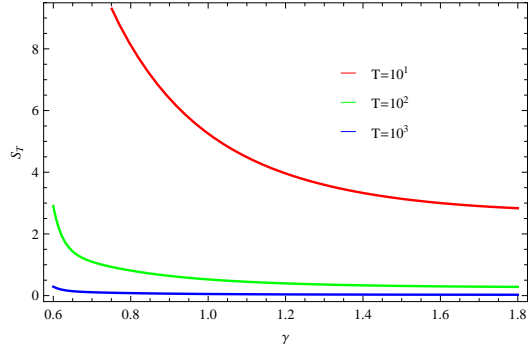,width=.55\linewidth}\caption{Plot of total
entropy versus $\gamma$.}
\end{figure}

\subsection{Thermal Equilibrium Scenario}

Further, we will discuss the thermal equilibrium scenario in the
present case. For thermodynamical equilibrium, the entropy function
attains a maximum value and satisfies the condition
$d^2S_T=d^2(S_{\textmd{A}}+S_{\textmd{f}})<0$. For this purpose, we
consider two cases of particle creation rate ($\Gamma$).

\subsection*{Case 1: \underline{$\Gamma$ = constant}}

Firstly, we consider particle creation rate $\Gamma$ as a constant.
Under this scenario, differentiating Eq.(\ref{S24}) w.r.t time, it
results the following second order differential equation
\begin{eqnarray}\nonumber
\ddot{S_T}&=&-\frac{3\gamma\big(1-\frac{\Gamma}{3H}\big){\big(-\frac{\epsilon}{r_c}+H\big)}\lambda
\dot{H}}{2H\big(-\frac{\epsilon}{r_c}+2H\big)^2}
+\frac{3\gamma\big(1-\frac{\Gamma}{3H}\big)\lambda\dot{H}}
{4H\big(-\frac{\epsilon}{r_c}+2H\big)}+\gamma\Gamma\big(1-\frac{\Gamma}{3H}\big)\nonumber\\
&\times&\frac{{\big(-\frac{\epsilon}{r_c}+H\big)}\lambda\dot{H}}
{4H^3\big(-\frac{\epsilon}{r_c}+2H\big)}
-\frac{3\gamma\big(1-\frac{\Gamma}{3H}\big){\big(-\frac{\epsilon}{r_c}+H\big)}
\lambda\dot{H}}{4H^2\big(-\frac{\epsilon}{r_c}+2H\big)}+\frac{3\gamma}
{4H\big(-\frac{\epsilon}{r_c}+2H\big)}\nonumber\\
&\times&\big(1-\frac{\Gamma}{3H}\big)\big(-\frac{\epsilon}{r_c}+H\big)
\bigg(\frac{4\big(-\frac{\epsilon}{r_c}+H^2\big)^{\frac{1-\gamma}{\gamma}}}
{{T_0}}\bigg(-1+\frac{\epsilon}{2r_cH}-\frac{3\gamma}{2H}\nonumber\\
&\times&\big(-\frac{\epsilon}{r_c}+H\big)\bigg)\dot{H}+\frac{4\big(1-\gamma\big)H}{\gamma
T_0}\big(-\frac{\epsilon
H}{r_c}+H^2\big)^{-1+\frac{1-\gamma}{\gamma}}\bigg(-1+\frac{\epsilon}{2Hr_c}\nonumber\\
&-&\frac{3\gamma\big(-\frac{\epsilon}{r_c}+H\big)}{2H}\bigg)
\big(-\frac{\epsilon}{r_c}+2H\dot{H}\big)+4H\big(-\frac{\epsilon
H}{r_c}+H^2\big)^\frac{1-\gamma}{\gamma}\nonumber\\
&\times&\frac{\big(-\frac{\epsilon\dot{H}}{2r_cH^2}-\frac{3\gamma\dot{H}}
{2H}+\frac{3\gamma\big(-\frac{\epsilon}{r_c}+H\big)}{2H^2}\big)}{T_0}\bigg)\label{S39}.
\end{eqnarray}
where $\lambda=1+\frac{4H}{T_0}\big(-\frac{\epsilon
H}{r_c}+H^2\big)^{\frac{1-\gamma}{\gamma}}\big(-1+\frac{\epsilon}{2r_c
H}-\frac{3\gamma(-\frac{\epsilon}{r_c}+H)}{2H}\big)$. The plot
between $\ddot{S_T}$ versus $\gamma$ for three values of $T$ with
constant values of $H=67$, $r_c=\frac{1}{67}$. $\epsilon=-1$,
$q=-0.53$ as shown in Figure {\bf2}. One can observe that the
thermodynamical equilibrium condition holds for all values of $T$
with specific ranges of $\gamma$. For example, for $T=10^2$,
thermodynamical equilibrium holds for the range $1.4<\gamma\leq1.8$
and does not obey for $0.6\leq\gamma\leq1.4$. For $T=10^{2.3}$,
thermal equilibrium holds for the range $1.3<\gamma\leq1.8$ and does
not showing the validity for $0.6\leq\gamma\leq1.3$. However, for
$T=10^{2.5}$, thermodynamic equilibrium condition holds for the
range $1.2<\gamma\leq1.8$ and disobey for the range
$0.6\leq\gamma<1.2$.
\begin{figure} \centering
\epsfig{file=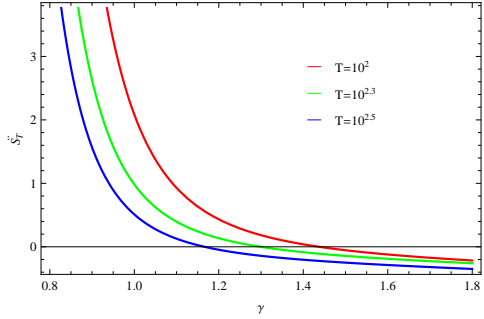,width=.50\linewidth}\caption{Plot of
$\ddot{S_T}$ versus $\gamma=constant$}
\end{figure}

\subsection*{Case 2: $\underline{\Gamma=\Gamma(t)}$}

Here we take $\Gamma$ as variable parameter, i.e.,
$\Gamma=\Gamma(t)$, for which Eq.(\ref{S24}) becomes
\begin{eqnarray}\nonumber
\ddot{S_T}&=&-\frac{3\gamma\lambda(-\frac{\epsilon}{r_c}+H)
(1-\frac{\Gamma}{3H})\dot{H}}{2H(-\frac{\epsilon}{r_c}+2H)^2}+\frac{3\gamma
\lambda(1-\frac{\Gamma}{3H})\dot{H}}{4H(-\frac{\epsilon}{r_c}+2H)}-3\gamma\lambda(1-\frac{\Gamma}{3H})\dot{H}\nonumber\\
&\times&\frac{
(-\frac{\epsilon}{r_c}+H)}{4H^2(-\frac{\epsilon}{r_c}+2H)}+\frac{3\gamma
(-\frac{\epsilon}{r_c}+H)
(1-\frac{\Gamma}{3H})}{4H(-\frac{\epsilon}{r_c}+2H)}\bigg(4(-\frac{\epsilon
H}{r_c}+H^2)^{\frac{1-\gamma}{\gamma}}\nonumber\\
&\times&\frac{\big(-1+\frac{\epsilon}{2r_c
H}-\frac{3\gamma(-\frac{\epsilon}{r_c}+H)}{2H}\big)\dot{H}}{T_0}+\frac{4\big(1-\gamma\big)H}{\gamma
T_0}(-\frac{\epsilon
H}{r_c}+H^2)^{-1+\frac{1-\gamma}{\gamma}}\nonumber\\
&\times&4H\big(1-\gamma\big)\big(-1+\frac{\epsilon}{2r_cH}-\frac{3\gamma
(-\frac{\epsilon}{r_c}+H)}{2H}\big)\big(-\frac{\epsilon\dot{H}}{r_c}+2H\dot{H}\big)\nonumber\\
&+&\frac{4H(-\frac{\epsilon
H}{r_c}+H^2)^{\frac{1-\gamma}{\gamma}}\big(-\frac{\epsilon\dot{H}}{2r_cH^2}-\frac{3\gamma\dot{H}}{2H}+\frac{3\gamma
(-\frac{\epsilon}{r_c}+H)\dot{H}}{2H^2}}{T_0}\bigg)\nonumber\\
&+&\frac{3\gamma\lambda
(-\frac{\epsilon}{r_c}+H)\big(\frac{\Gamma\dot{H}}{3H^2}-\frac{\dot{\Gamma}}{3H}\big)}{4H(-\frac{\epsilon}{r_c}+2H)},
\end{eqnarray}
where
\begin{eqnarray}\nonumber
\dot{\Gamma}=-\frac{9\gamma\big(\frac{\Gamma}{3H}\big)\big(H-\frac{\epsilon}{r_c}\big)\big(1-\frac{\Gamma}{3H}\big)}
{2H-\frac{\epsilon}{r_c}}+3H\Gamma\big(1-\frac{\Gamma}{3H}\big).
\end{eqnarray}
The plot of $\ddot{S_T}$ versus $\gamma$ for three values of $T$  as
shown in Figure \textbf{3} by keeping the same constant values as in
previous case. We observe that the thermodynamic equilibrium holds
for all values of $T$ with different ranges of $\gamma$. For
example, for $T=10^2$, thermodynamic equilibrium holds for the range
$1.4<\gamma\leq1.8$ and does not satisfying $0.6\leq\gamma\leq1.4$.
For $T=10^{2.3}$, it leads to the validity of thermodynamic
equilibrium for the range $1.3<\gamma\leq1.8$ and does not valid for
$0.6\leq\gamma\leq1.3$. However, for $T=10^{2.5}$, thermodynamic
equilibrium holds for the range $1.2\leq\gamma\leq1.8$ and does not
satisfying within $0.6\leq\gamma<1.2$.
\begin{figure} \centering
\epsfig{file=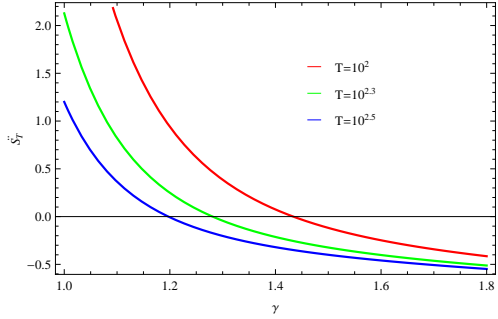,width=.50\linewidth}\caption{Plot of
$\ddot{S_T}$ versus $\gamma=\Gamma(t)$}
\end{figure}

\section{Logarithmic Corrected Entropy}

Quantum gravity allows the logarithmic corrections in the presence
of thermal equilibrium fluctuations and quantum fluctuations
\cite{15}-\cite{21}. The logarithmic entropy corrections can be
defined as
\begin{eqnarray}\label{S25}
S_{\textmd{A}}=\frac{A}{4L^2_{\textmd{p}}}+\alpha\ln\frac{A}{4L^2_{\textmd{p}}}+\beta\frac{4L^2_{\textmd{p}}}{A},
\end{eqnarray}
where $\alpha$ and $\beta$ are constants whose  values are still
under consideration. The differential form of above equation leads
to
\begin{eqnarray}\label{S28}
dS_{\textmd{A}}=-\frac{3\gamma\big(H-\frac{\epsilon}{r_{\textmd{c}}}\big)
\big(1-\frac{\Gamma}{3H}\big)}{\big(2H-\frac{\epsilon}{r_{\textmd{c}}}\big)}
\bigg(\frac{1}{4HL^2_{\textmd{p}}}+2\alpha H-16\beta
H^3L^2_{\textmd{p}}\bigg)dt,
\end{eqnarray}
which gives
\begin{eqnarray}\label{S26}
T_{\textmd{A}}dS_{\textmd{A}}=\frac{3\gamma\big(H-\frac{\epsilon}{r_{\textmd{c}}}\big)\big(1-\frac{\Gamma}{3H}\big)}
{\big(2H-\frac{\epsilon}{r_{\textmd{c}}}\big)}
\bigg(\frac{1}{L^2_{\textmd{p}}}+8H^2\alpha-64\beta
H^4L^2_{\textmd{p}}\bigg)dt.
\end{eqnarray}
In view of this entropy, the quantity $\Xi$ takes the form
\begin{eqnarray}\label{S44}
\Xi=3\gamma(H-\frac{\epsilon}{r_c})\bigg(-\frac{1}{2H}+\frac{(1-\frac{\Gamma}{3H})}
{(2H-\frac{\epsilon}{r_c})}\big(\frac{1}{L^2_{\textmd{p}}}+8H^2\alpha-64\beta
H^4L^2_{\textmd{p}}\big)\bigg).
\end{eqnarray}
It can be observed from Eq.(\ref{S44}) that the first law of
thermodynamics holds when
$\Gamma=3H\big(1-\frac{(2H-\frac{\epsilon}{r_c})}{2H}(\frac{1}{L^2_{\textmd{p}}}+8H^2\alpha-64\beta
H^4L^2_{\textmd{p}})\big)$. To discuss the GSLT for logarithmic
corrected entropy of horizon, we obtain the total entropy by using
Eqs.(\ref{S23}) and (\ref{S28}) as follows
\begin{eqnarray}\nonumber
\dot{S_T}&=&\frac{3\gamma\big(H-\frac{\epsilon}{r_{\textmd{c}}}\big)
\big(1-\frac{\Gamma}{3H}\big)}{\big(2H-\frac{\epsilon}{r_{\textmd{c}}}\big)}
\bigg(4HL^2_{\textmd{p}}+2\alpha H-16\beta H^3L^2_{\textmd{p}}
-T^{-1}_0\\\label{S29}&\times&
\big(H^2-\frac{\epsilon}{r_{\textmd{c}}}H\big)^\frac{1-\gamma}{\gamma}
\bigg(1-\frac{1}{2H}\frac{\epsilon}{r_{\textmd{c}}}
-\frac{3\gamma}{2H}\big(H-\frac{\epsilon}{r_{\textmd{c}}}\big)\bigg)\bigg)\label{S40}.
\end{eqnarray}
The GSLT will hold under these constraints.
\begin{itemize}
  \item For the case \underline{$\Gamma< 3H$}, the GSLT satisfy in the quintessence region of the universe if the following
  constraint holds
  \begin{eqnarray}\nonumber
4L^2_{\textmd{p}}+2\alpha&>&16\beta H^2L^2_{\textmd{p}} +(HT_0)^{-1}
\bigg(1-\frac{1}{2H}\frac{\epsilon}{r_{\textmd{c}}}
-\frac{3\gamma}{2H}\big(H-\frac{\epsilon}{r_{\textmd{c}}}\big)\bigg)
\\\nonumber&\times&
\big(H^2-\frac{\epsilon}{r_{\textmd{c}}}H\big)^\frac{1-\gamma}{\gamma}.
\end{eqnarray}
  \item \underline{$\Gamma> 3H$} For this case, we obtain the
  following
  constraint
  \begin{eqnarray}\nonumber
4L^2_{\textmd{p}}+2\alpha&<&16\beta H^2L^2_{\textmd{p}} +(HT_0)^{-1}
\bigg(1-\frac{1}{2H}\frac{\epsilon}{r_{\textmd{c}}}
-\frac{3\gamma}{2H}\big(H-\frac{\epsilon}{r_{\textmd{c}}}\big)\bigg)
\\\nonumber&\times&
\big(H^2-\frac{\epsilon}{r_{\textmd{c}}}H\big)^\frac{1-\gamma}{\gamma}.
\end{eqnarray}
which implies the GSLT holds in phantom era of the universe.
  \item The case \underline{$\Gamma= 3H$} means $\dot{S}_{T}=0$ in the cosmological constant era.
\end{itemize}

The expression of total entropy in the form of Hubble parameter is
given by
\begin{eqnarray}\label{S33}
S_{\textmd{T}}&=&S_{\textmd{A}}+S_{\textmd{f}}\nonumber\\
&=&-8H^2\beta
L^2_{\textmd{p}}+2\ln(H)(\alpha+2L^2_{\textmd{p}})-\frac{\gamma\big(H(H-\frac{\epsilon}{r_c}
)\big)^{\frac{1}{\gamma}}}{2H^3T_0}.
\end{eqnarray}
The plot of total entropy $S_T$ versus $\gamma$ with respect to
three values of $T$ is shown in Figure \textbf{4} with constant
values as $\alpha=-2$, $\beta=-0.00001$, $L_p=1$. It is observed
that the total entropy is positive, i.e, $S_T>0$ which leads to the
validity of GSLT for all values of $T$ .

\subsection{Thermal Equilibrium Scenario}

Now we will discuss the thermodynamic equilibrium by assuming two
cases for particle creation rate $\Gamma$ as follows:
\begin{figure} \centering
\epsfig{file=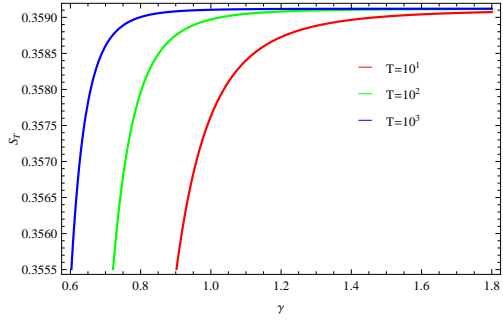,width=.50\linewidth}\caption{Plots of total
entropy versus $\gamma$}
\end{figure}
\subsection*{Case 1: \underline{$\Gamma$ = constant}}

In this way, the second order differential equation can be obtained
from Eq.(\ref{S40}) for $\Gamma$ as a constant
\begin{eqnarray}\nonumber
\ddot{S_T}&=&-\frac{6\gamma
(1-\frac{\Gamma}{3H})\big(H-\frac{\epsilon}{r_c}\big)}{\big(2H-\frac{\epsilon}{r_c}\big)}
\big(4HL^2_{\textmd{p}}-\lambda_2+2\alpha
H-16L^2_{\textmd{p}}\beta H^3\big)\dot{H}+3\gamma\nonumber\\
&\times&\frac{
(1-\frac{\Gamma}{3H})\big(4HL^2_{\textmd{p}}-\lambda_2+2\alpha
H-16L^2_{\textmd{p}}\beta
H^3\big)\dot{H}}{\big(2H-\frac{\epsilon}{r_c}\big)}+\gamma\Gamma\big(H-\frac{\epsilon}{r_c}\big)\nonumber\\
&\times&\frac{\big(4HL^2_{\textmd{p}}-\lambda_2+2\alpha
H-16L^2_{\textmd{p}}\beta
H^3\big)}{H^2\big(2H-\frac{\epsilon}{r_c}\big)}+\frac{3\gamma
(1-\frac{\Gamma}{3H})\big(H-\frac{\epsilon}{r_c}\big)}{\big(2H-\frac{\epsilon}{r_c}\big)}\nonumber\\
&\times&\bigg(4L^2_{\textmd{p}}\dot{H}-\frac{\big(1-\gamma\big)
\big(1-\frac{3\gamma\big(H-\frac{\epsilon}{r_c}\big)}{2H}-\frac{\epsilon}{2Hr_c}\big)\big(H^2-\frac{\epsilon
H}{r_c}\big)^{-1+\frac{1-\gamma}{\gamma}}}{\gamma T_0}\nonumber\\
&\times&\big(2H\dot{H}-\frac{\epsilon\dot{H}}{r_c}\big)
-\frac{\big(H^2-\frac{\epsilon}{r_c}\big)^{\frac{1-\gamma}{\gamma}}\big(-\frac{3\gamma\dot{H}}{2H}+
\frac{3\gamma\big(H-\frac{\epsilon}{r_c}\big)}{2H^2}+\frac{\epsilon\dot{H}}{2H^2r_c}\big)}{T_0}\nonumber\\
&+&2\alpha\dot{H}-48L^2_{\textmd{p}}\beta H^2\beta\dot{H}\bigg),
\end{eqnarray}
where
\begin{eqnarray}\nonumber
\lambda_2=\frac{\big(1-\frac{3\gamma\big(H-\frac{\epsilon}{r_c}\big)}
{2H}-\frac{\epsilon}{2Hr_c}\big)\big(H^2-\frac{\epsilon
H}{r_c}\big)^{\frac{1-\gamma}{\gamma}}}{T_0}\nonumber.
\end{eqnarray}
The plot between $\ddot{S_T}$ and $\gamma$ for three values of $T$
by fixing the constant values $q=-0.53$,
$\alpha=-2,~\beta=-0.000000001$ and $L_p=1$ as shown in Figure
\textbf{5}. It can be seen that thermodynamical equilibrium is
obeying the condition $\ddot{S_T}<0$ for all values of $T$ which
leads to the thermal equilibrium.
\begin{figure} \centering
\epsfig{file=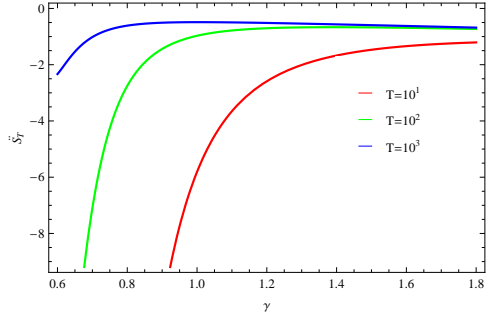,width=.50\linewidth}\caption{Plot of
$\ddot{S_T}$ versus $\gamma=constant$}
\end{figure}

\subsection*{Case 2: $\underline{\Gamma=\Gamma(t)}$}

Taking $\Gamma$ as a function of $t$, Eq.(\ref{S40}) yields
\begin{eqnarray}\nonumber
\ddot{S_T}&=&-\frac{6\gamma
(1-\frac{\Gamma}{3H})\big(H-\frac{\epsilon}{r_c}\big)}{\big(2H-\frac{\epsilon}{r_c}\big)}
\big(4HL^2_{\textmd{p}}-\lambda_2+2\alpha
H-16L^2_{\textmd{p}}\beta H^3\big)\dot{H}+3\gamma\nonumber\\
&\times&\frac{
(1-\frac{\Gamma}{3H})\big(4HL^2_{\textmd{p}}-\lambda_2+2\alpha
H-16L^2_{\textmd{p}}\beta
H^3\big)\dot{H}}{\big(2H-\frac{\epsilon}{r_c}\big)}+\frac{3\gamma
(1-\frac{\Gamma}{3H})}{\big(2H-\frac{\epsilon}{r_c}\big)}\nonumber\\
&\times&\big(H-\frac{\epsilon}{r_c}\big)\bigg(4L^2_{\textmd{p}}\dot{H}-\frac{
\big(1-\frac{3\gamma\big(H-\frac{\epsilon}{r_c}\big)}{2H}-\frac{\epsilon}{2Hr_c}\big)\big(H^2-\frac{\epsilon
H}{r_c}\big)^{-1+\frac{1-\gamma}{\gamma}}}{\gamma T_0}\nonumber\\
&\times&\big(1-\gamma\big)\big(2H\dot{H}-\frac{\epsilon\dot{H}}{r_c}\big)-
\frac{\big(-\frac{3\gamma\dot{H}}{2H}+
\frac{3\gamma\big(H-\frac{\epsilon}{r_c}\big)}{2H^2}+\frac{\epsilon\dot{H}}{2H^2r_c}\big)}{T_0}\nonumber\\
&\times&\big(H^2-\frac{\epsilon}{r_c}\big)^{\frac{1-\gamma}{\gamma}}+2\alpha\dot{H}-48L^2_{\textmd{p}}\beta
H^2\beta\dot{H}\bigg)+3\gamma\big(H-\frac{\epsilon}{r_c}\big)
\nonumber\\
&\times&\frac{1}{\big(2H-\frac{\epsilon}{r_c}\big)}\big(4HL^2_{\textmd{p}}-\lambda_2+2\alpha
H-16L^2_{\textmd{p}}\beta
H^3\big)\big(\frac{\Gamma\dot{H}}{3H^2}-\frac{\dot{\Gamma}}{3H}\big).
\end{eqnarray}
Figure \textbf{6} reperesents the plot between $\ddot{S_T}$ and
$\gamma$ for three values of $T$ for variable $\Gamma$ for same
constant values. Figure \textbf{6} indicates that the trajectories
of $\ddot{S_T}$ corresponding to all the values of $T$ ensure the
validity of thermodynamical equilibrium.
\begin{figure} \centering
\epsfig{file=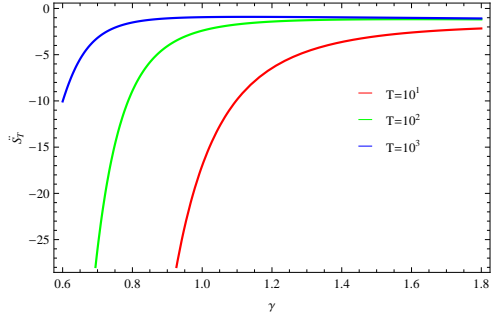,width=.50\linewidth}\caption{Plot of
$\ddot{S_T}$ versus $\gamma=\Gamma(t)$}
\end{figure}

\section{Power Law Corrected Entropy}

The power-law correction to the entropy-area law comes from
association of the wave-function of the scalar field between the
ground state and the exited state \cite{22}-\cite{24}. The
correction term is also more significant for higher excitations. It
is important to note that the correction term decreases faster with
A and hence in the semi-classical limit (large area) the
entropy-area law is recovered. The power entropy can b given as
\begin{eqnarray}\label{S30}
S_{\textmd{A}}=\frac{A}{4L^2_{\textmd{p}}}\big(1-K_\delta
A^{1-\frac{\delta}{2}}\big),~~~K_\delta=\frac{\delta\big(4\pi\big)^{\frac{\delta}{2}-1}}{\big(4-\delta\big)r^{4-\delta}_c}
\end{eqnarray}
where $ \delta $ is dimensionless constant and $r_c$ is the
crossover scale. From Eq.(\ref{S30}) the differential of surface
entropy at horizon can be expressed as
\begin{eqnarray}\label{S41}
dS_{\textmd{A}}=\frac{3\gamma\big(H-\frac{\epsilon}{r_c}\big)
\big(1-\frac{\Gamma}{3H}\big)}{\big(2H-\frac{\epsilon}{r_c}\big)}
\bigg(\frac{1}{4HL^2_{\textmd{p}}}-\frac{K_\delta}{4HL^2_{\textmd{p}}}
\big(2-\frac{\delta}{2}\big)\big(\frac{1}{H}\big)^{3-\delta}\bigg)dt,
\end{eqnarray}
which gives
\begin{eqnarray}\label{S31}
T_{\textmd{A}}dS_{\textmd{A}}=\frac{3\gamma\big(H-\frac{\epsilon}{r_c}\big)
\big(1-\frac{\Gamma}{3H}\big)}{\big(2H-\frac{\epsilon}{r_c}\big)}
\bigg(\frac{1}{L^2_{\textmd{p}}}-\big(2-\frac{\delta}{2}\big)K_\delta
R^{2-\delta}\bigg)dt.
\end{eqnarray}
In this entropy correction, we have
\begin{eqnarray}\label{S46}
\Xi=3\gamma(H-\frac{\epsilon}{r_c})\bigg(-\frac{1}{2H}+\frac{(1-\frac{\Gamma}{3H})}{(2H-\frac{\epsilon}{r_c}}
\big(\frac{1}{L^2_{\textmd{p}}}-(2-\frac{\delta}{2})K_\delta
R^{2-\delta}\big)\bigg),
\end{eqnarray}
This equation shows that the first law of thermodynamics (i.e.,
$\Xi=0$) holds for
$\Gamma=3H\big(1-\frac{(2H-\frac{\epsilon}{r_c})}{2H\big(\frac{1}{L^2_{\textmd{p}}}-(2-\frac{\delta}{2})K_\delta
R^{2-\delta}\big)}\big)$.

In the presence of power law corrected entropy, the rate of change
of total entropy takes the form
\begin{eqnarray}\nonumber
\dot{S_T}=\frac{3\gamma\big(H-\frac{\epsilon}{r_c}\big)\big(1-\frac{\Gamma}{3H}\big)}{\big(2H-\frac{\epsilon}{r_c}\big)}
\bigg(\frac{1}{4HL^2_{\textmd{p}}}-\frac{K_\delta}{4HL^2_{\textmd{p}}}
\big(2-\frac{\delta}{2}\big)\big(\frac{1}{H}\big)^{3-\delta}
\\-T^{-1}_0\big(H^2-\frac{\epsilon}{r_c}H\big)^\frac{1-\gamma}{r_c}
\big(1-\frac{1}{2H}\frac{\epsilon}{r_c}-\frac{3\gamma}
{2H}(H-\frac{\epsilon}{r})\big)\bigg)\label{S32}.
\end{eqnarray}
To discuss the validity of GSLT, the following constraints must
satisfy. These are as
\begin{itemize}
  \item The condition \underline{$\Gamma<3H$} implies that
  \begin{eqnarray}\nonumber
\frac{1}{4L^2_{\textmd{p}}}&>&\frac{K_\delta}{4L^2_{\textmd{p}}}
\big(2-\frac{\delta}{2}\big)\big(\frac{1}{H}\big)^{3-\delta}
+HT^{-1}_0 \big[1-\frac{1}{2H}\frac{\epsilon}{r_c}-\frac{3\gamma}
{2H}(H-\frac{\epsilon}{r})\big]\\\nonumber&\times&\big(H^2-\frac{\epsilon}{r_c}H\big)^\frac{1-\gamma}{r_c}
\end{eqnarray}
\item \underline{$\Gamma>3H$}
For this case, we obtain the following constraint which indicates
the validity of GSLT in phantom phase. It is given as
  \begin{eqnarray}\nonumber
\frac{1}{4L^2_{\textmd{p}}}&<&\frac{K_\delta}{4L^2_{\textmd{p}}}
\big(2-\frac{\delta}{2}\big)\big(\frac{1}{H}\big)^{3-\delta}
+HT^{-1}_0 \big[1-\frac{1}{2H}\frac{\epsilon}{r_c}-\frac{3\gamma}
{2H}(H-\frac{\epsilon}{r})\big]\\\nonumber&\times&\big(H^2-\frac{\epsilon}{r_c}H\big)^\frac{1-\gamma}{r_c}
\end{eqnarray}
  \item The case \underline{$\Gamma= 3H$} gives $\dot{S}_{T}=0$ in the cosmological constant era.
\end{itemize}
\begin{figure} \centering
\epsfig{file=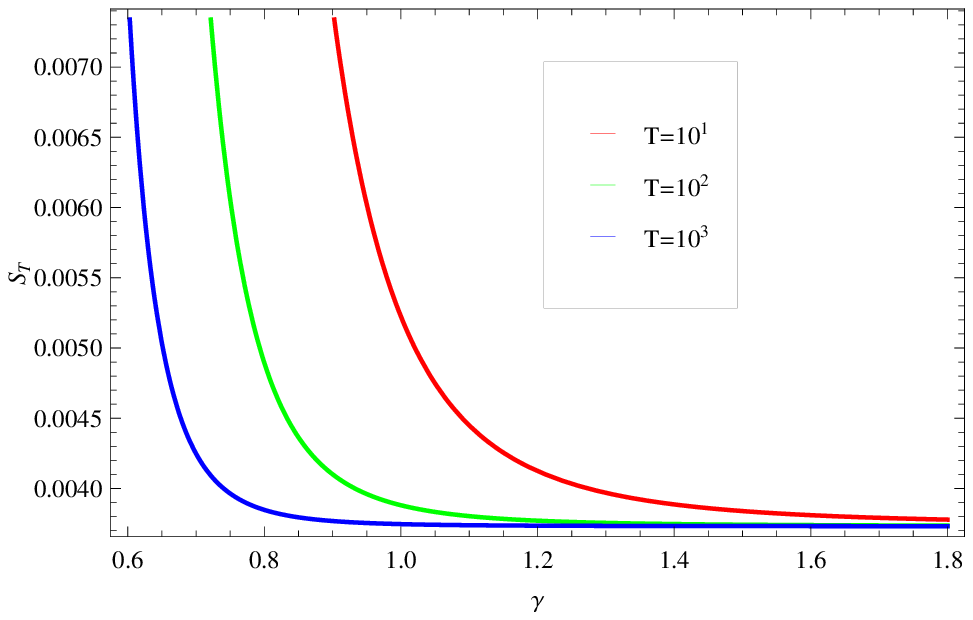,width=.50\linewidth}\caption{Plot of total
entropy versus $\gamma$}
\end{figure}
The equation (\ref{S32}) in terms of Hubble parameter takes the
following form
\begin{eqnarray}\label{34}
S_{T}&=&S_{\textmd{A}}+S_{\textmd{f}}\nonumber\\
&=&\frac{1}{4}\bigg(\frac{1}{HL^2_{\textmd{p}}}
-\frac{(\frac{1}{H})^{5-\delta}(-4+\delta)
K_\delta}{2L^2_{\textmd{p}}(-5+\delta)} +\frac{2\gamma(H(H
-\frac{\epsilon}{r_c}))^{\frac{1}{\gamma}}}{H^3T_0}\bigg).
\end{eqnarray}
The plot of total entropy $(S_T)$ is shown in Figure \textbf{7}
versus $\gamma$ for three values of $T$ by assigning the constant
values as $L = 1$, $H=67$, $\frac{1}{67}$, $\alpha = -2$, and
$\epsilon = -1$. We observe that total entropy is positive, i.e,
$(S_T)>0 $ which leads to the validity of GSLT for all values of
$T$.

\subsection{Thermal Equilibrium Scenario}

For thermodynamic equilibrium, we assume two cases of $\Gamma$,
i.e., $\Gamma$ is constant and  $\Gamma$ is variable.

\subsection*{Case 1: \underline{$\Gamma$ = constant}}
For constant $\Gamma$, the second order differential equation of
total entropy leads to
\begin{eqnarray}\nonumber
\ddot{S_T}&=&-\frac{6\gamma
(1-\frac{\Gamma}{3H})(H-\frac{\epsilon}{r_c})
\lambda_3}{(2H-\frac{\epsilon}{r_c})^2}+\frac{3\gamma
(1-\frac{\Gamma}{3H})\lambda_3}{(2H-\frac{\epsilon}{r_c})}+\frac{\gamma\Gamma
(H-\frac{\epsilon}{r_c})\lambda_3}{H^2
(2H-\frac{\epsilon}{r_c})}\nonumber\\
&+&\frac{3\gamma(1-\frac{\Gamma}{3H})(H-\frac{\epsilon}{r_c})}{(2H-\frac{\epsilon}{r_c})}
\bigg(-\frac{\dot{H}}{4H^2L^2_{\textmd{p}}}
+\frac{\big(4-\delta\big)\big(2-\frac{\delta}{2}\big)\big(\frac{1}{4H}\big)^{5-\delta}
K_\delta\dot{H}}{4L^2_{\textmd{p}}}\nonumber\\
&-&\frac{\big(1-\gamma\big)\big(1-\frac{3\gamma\big(H-\frac{\epsilon}{r_c}\big)}{2H}
-\frac{\epsilon}{2Hr_c}\big)\big(H^2-\frac{\epsilon
H}{r_c}\big)^{-1+\frac{1-\gamma}{\gamma}}\big(2H\dot{H}-\frac{\epsilon\dot{H}}{r_c}\big)}{\gamma
T_0}\nonumber\\
&-&\frac{\big(H^2-\frac{\epsilon
H}{r_c}\big)^{\frac{1-\gamma}{\gamma}}\big(-\frac{3\gamma\dot{H}}{2H}+\frac{3\gamma\big(H-\frac{\epsilon}{r_c}\big)\dot{H}}{2H^2}+
\frac{\epsilon\dot{H}}{r_c}\big)}{T_0}\bigg),
\end{eqnarray}
where
\begin{eqnarray}\nonumber
\lambda_3&=&\bigg(\frac{1}{4HL^2_{\textmd{p}}}-\frac{\big(2-\frac{\delta}{2}\big)\big(\frac{1}{H}\big)^{4-\delta}K_\delta}
{4L^2_{\textmd{p}}}-\big(1-\frac{3\gamma\big(H-\frac{\epsilon}{r_c}\big)}{2H}-\frac{\epsilon}{2Hr_c}\big)\nonumber\\
&\times&\frac{\big(H^2-\frac{\epsilon
H}{r_c}\big)^{\frac{1-\gamma}{\gamma}}}{T_0}\bigg)\dot{H}\nonumber.
\end{eqnarray}
The graphical behavior of $\ddot{S_T}$ against $\gamma$ is shown in
Figure \textbf{8} for three values of $T$ by keeping the constant
values as $H=67$, $r_c=\frac{1}{67}$, $\epsilon=-1$, $q=-0.53$,
$\delta=-2$, $L=1$. It is found that the trajectories of this plot
satisfy the condition $\ddot{S_T}<0$ for all values of $T$ which
leads to the thermodynamical equilibrium of the system.
\begin{figure} \centering
\epsfig{file=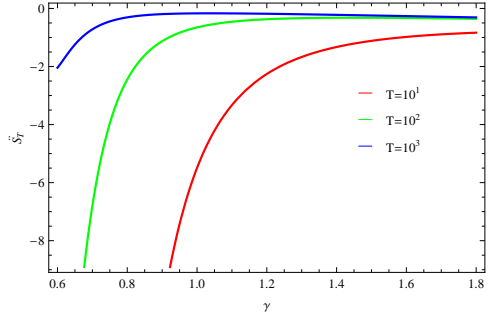,width=.50\linewidth}\caption{Plot of
$\ddot{S_T}$ versus $\gamma$}
\end{figure}
\subsection*{Case 2: $\underline{\Gamma=\Gamma(t)}$}

For variable $\Gamma$, second order differential equation of total
entropy turns out to be
\begin{eqnarray}\nonumber
\ddot{S_T}&=&-\frac{6\gamma
(1-\frac{\Gamma}{3H})(H-\frac{\epsilon}{r_c})
\lambda_3}{(2H-\frac{\epsilon}{r_c})^2}+\frac{3\gamma
(1-\frac{\Gamma}{3H})\lambda_3}{(2H-\frac{\epsilon}{r_c})}+\frac{3\gamma
(1-\frac{\Gamma}{3H})
(H-\frac{\epsilon}{r_c})}{(2H-\frac{\epsilon}{r_c})}\nonumber\\
&\times&\bigg(-\frac{\dot{H}}{4H^2L^2_{\textmd{p}}}
+\frac{\big(4-\delta\big)\big(2-\frac{\delta}{2}\big)\big(\frac{1}{4H}\big)^{5-\delta}K_\delta\dot{H}}
{4L^2_{\textmd{p}}}-\big(2H\dot{H}-\frac{\epsilon\dot{H}}{r_c}\big)\nonumber\\
&\times&\frac{(1-\gamma)\big(1-\frac{3\gamma\big(H-\frac{\epsilon}{r_c}\big)}
{2H}-\frac{\epsilon}{2Hr_c}\big)\big(H^2-\frac{\epsilon
H}{r_c}\big)^{-1+\frac{1-\gamma}{\gamma}}}{\gamma
T_0}-\big(H^2-\frac{\epsilon
H}{r_c}\big)^{\frac{1-\gamma}{\gamma}}\nonumber\\
&\times&\frac{\big(-\frac{3\gamma\dot{H}}{2H}+\frac{3\gamma\big(H-\frac{\epsilon}{r_c}\big)\dot{H}}{2H^2}+
\frac{\epsilon\dot{H}}{2H^2r_c}\big)}{T_0}\bigg)
+\frac{3\gamma-\big(\frac{\epsilon}{r_c}\big)\lambda_3\big(\frac{\Gamma
\dot{H}}{3H^2}-\frac{\dot{\Gamma}}{3H}\big)}{2H-\frac{\epsilon}{r_c}}.
\end{eqnarray}
The plot between $\ddot{S_T}$ and $\gamma$ for three values of $T$
by setting the constant values as $H=67$, $r_c=\frac{1}{67}$,
$\epsilon=-1$, $q=-0.53$, $\delta=-2$, $L=1$ as shown in Figure
\textbf{9}. We observe that thermodynamical equilibrium satisfying
the condition $\ddot{S_T}<0$ for all values of $T$ which leads to
thermodynamical disequilibrium of the system.
\begin{figure} \centering
\epsfig{file=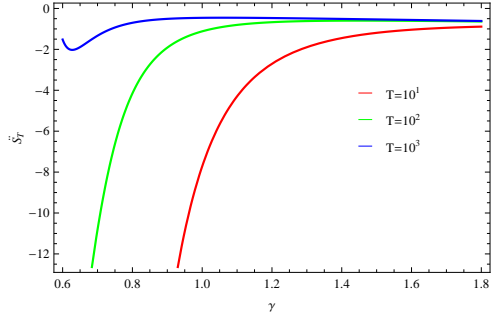,width=.50\linewidth}\caption{Plot of
$\ddot{S_T}$ versus $\gamma$}
\end{figure}

\section{Comparison and Conclusion}

We have studied the thermodynamics on the apparent horizon for
gravitationally induced particle creation scenario by assuming
entropy corrections (Bakenstein entropy, logarithmic corrected
entropy and power law corrected entropy) in DGP braneworld gravity.
Considering the perfect fluid EoS $p=(\gamma-1)\rho$, we have
analyzed the first law of thermodynamics, GSLT and thermodynamic
equilibrium. We have fixed model parameters in the way to obtain the
validity of GSLT. However, the results may be changed for other
model parameters. The results have been summarized as follows:

\underline{For Bakenstein entropy}, the first law of thermodynamics
holds at apparent horizon when
$\Gamma=3H\big(1-\frac{(2H-\frac{\epsilon}{r_c})}{2H}\big)$. The
GSLT valid under some constraints in different eras of the evolving
universe. In terms of total entropy scenario, GSLT remains valid for
all values of $T$ with $\frac{2}{3}\leq\gamma\leq 2$ (Figure
\textbf{1}). In the presence of usual entropy of horizon and for
constant $\Gamma$, thermodynamical equilibrium holds for all values
of $T$ with specific ranges of $\gamma$. For example,
thermodynamical equilibrium is satisfying the condition
$\frac{d^2S_T}{dt^2}<0$ for $1.4<\gamma\leq1.8$, $3<\gamma\leq1.8$,
$1.2<\gamma\leq1.8$ and does not showing the validity
$0.6\leq\gamma\leq1.4$, $0.6\leq\gamma\leq1.3$ and
$0.6\leq\gamma<1.2$ with $T=10^2$, $T=10^{2.3}$ and $T=10^{2.5}$
respectively (Figure \textbf{2}). For variable $\Gamma$, Figure
\textbf{3} indicates that thermodynamical equilibrium holds for all
values of $T$ with different ranges of $\gamma$ such as
$1.4<\gamma\leq1.8$, $1.3<\gamma\leq1.8$, $1.2\leq\gamma\leq1.8$ and
does not obey $0.6\leq\gamma\leq1.4$, $0.6\leq\gamma\leq1.3$ and
$0.6\leq\gamma<1.2$ for $T=10^2$, $T=10^{2.3}$ and $T=10^{2.5}$
respectively.

\underline{For logarithmic corrected entropy}, it has observed that
first law of thermodynamics holds at apparent horizon when
$\Gamma=3H\big(1-\frac{(2H-\frac{\epsilon}{r_c})}{2H}(\frac{1}{L^2_{\textmd{p}}}+8H^2\alpha-64\beta
H^4L^2_{\textmd{p}})\big)$. The GSLT valid under some constraints in
different eras of the evolving universe. In terms of total entropy
scenario, GSLT is valid for all values of $T$ with
$\frac{2}{3}\leq\gamma\leq 2$ (Figure \textbf{4}). Thermodynamical
equilibrium condition ($\frac{d^2S_T}{dt^2}<0$) satisfied for all
values of $T$ with all values of $\gamma$ for constant as well as
variable $\Gamma$ (Figure \textbf{5} and \textbf{6}).

\underline{For power law entropy corrected}, it has been observed
that the first law of thermodynamics satisfies under constraint
$\Gamma=3H\big(1-\frac{(2H-\frac{\epsilon}{r_c})}{2H\big(\frac{1}{L^2_{\textmd{p}}}-(2-\frac{\delta}{2})K_\delta
R^{2-\delta}\big)}\big)$ and GSLT holds under some conditions on
model parameters in quintessence, phantom and cosmological constant
phases. Also, GSLT meets for all values of $T$ in the range
$(\frac{2}{3}\leq\gamma\leq 2)$ (Figure \textbf{7}). It is found
that, for constant as well as variable $\Gamma$, the thermodynamic
equilibrium condition $\frac{d^2S_T}{dt^2}<0$ obey for all values of
$T$ which leads to the thermodynamical equilibrium of the system.

Here we provide some details about past works and compare with
underlying work. Harko et al. \cite{harko} considered the
possibility of a gravitationally induced particle production through
the mechanism of a non-minimal curvature–matter coupling. An
interesting feature of this gravitational theory is that the
divergence of the energy– momentum tensor is nonzero. Firstly, they
have reformulated the model in terms of an equivalent scalar–tensor
theory, with two arbitrary potentials. By using the formalism of
open thermodynamic systems, they have interpreted the energy balance
equations in this gravitational theory from a thermodynamic point of
view, as describing irreversible matter creation processes. The
particle number creation rates, the creation pressure, and the
entropy production rates have explicitly obtained as functions of
the scalar field and its potentials, as well as of the matter
Lagrangian. The temperature evolution laws of the newly created
particles are also obtained. The cosmological implications of the
model have briefly investigated, and it is shown that the late-time
cosmic acceleration may be due to particle creation processes.
Furthermore, it has also shown that due to the curvature–matter
coupling, during the cosmological evolution a large amount of
comoving entropy is also produced.

Mitra et al. \cite{R3} have studied thermodynamics laws by assuming
flat FRW universe enveloped by by apparent and event horizon in the
framework of RSII brane model and DGP brane scenario. Assuming
extended Hawking temperature on the horizon, the unified first law
is examined for perfect fluid (with constant equation of state) and
Modified Chaplygin Gas model. As a result there is a modification of
Bekenstein entropy on the horizons. Further the validity of the
generalized second law of thermodynamics and thermodynamical
equilibrium are also investigated.

Pan et al. \cite{pan} investigated the expansion of the universe
powered by the gravitationally induced adiabatic matter creation by
developing general creation rate and their dynamical analysis. They
also developed dynamical analysis in the the presence of a
non-singular universe (without the big bang singularity) with two
successive accelerated phases, one at the very early phase of the
universe (i.e. inflation), and the other one describes the current
accelerating universe, where this early, late accelerated phases are
associated with an unstable fixed point (i.e. repeller) and a stable
fixed point (attractor), respectively.

Saló and Haro \cite{salo} performed a qualitative and thermodynamic
study of two models when one takes into account adiabatic particle
production. In the first one, there is a constant particle
production rate, which leads to solutions depicting the current
cosmic acceleration but without inflation. The other one has
solutions that unify the early and late time acceleration. These
solutions converge asymptotically to the thermal equilibrium.

Recently, by assuming the gravitationally induced particle scenario
with constant specific entropy and arbitrary particle creation rate
($\Gamma$), thermodynamics on the apparent horizon for FRW universe
has been discussed \cite{1j}. They have investigated the first law,
GSLT and thermodynamical equilibrium by assuming the EoS for perfect
fluid and put forward various constraints on $\Gamma$ for which
thermodynamical laws hold. We have extended the work of \cite{1j} in
the DGP brane-world scenario by assuming usual entropy as well as
its entropy corrections (power law as well as logarithmic corrected)
in flat FRW universe. We have extracted EoS parameter and obtained
its various constraints with respect to quintessence, vacuum and
phantom era of the universe. For variable as well as constant
particle creation rate ($\Gamma$), the first law of thermodynamics,
GSLT and thermal equilibrium condition
is satisfied in all cases of entropies forms.\\

\end{document}